\documentclass[pra,twocolumn]{revtex4}
\usepackage{epsfig,amsmath,amssymb,mathtools}
\usepackage[active]{srcltx}
\usepackage[hypertex,linkcolor=red]{hyperref}
\usepackage{yfonts}

\def\R{\textfrak{R}}
\def\a{\alpha}
\def\b{\beta}
\begin{document}
\title{Emergence of Space-Time from Topologically Homogeneous Causal Networks}
\author{Giacomo Mauro D'Ariano} \affiliation{QUIT Group, Dipartimento di
  Fisica ``A. Volta'', via Bassi 6, I-27100 Pavia, Italy and Istituto
  Nazionale di Fisica Teorica e Nucleare, Sezione di Pavia.}
\author{Alessandro Tosini} \affiliation{QUIT Group, Dipartimento di Fisica
  ``A. Volta'', via Bassi 6, I-27100 Pavia, Italy and Istituto
  Nazionale di Fisica Teorica e Nucleare, Sezione di Pavia.}
\begin{abstract}
  In this paper we study the emergence of Minkowski space-time from a causal network. Differently
  from previous approaches, we require the network to be topologically homogeneous, so that the
  metric is derived from pure event-counting. Emergence from events has an operational motivation in
  requiring that every physical quantity---including space-time---be defined through precise
  measurement procedures. Topological homogeneity is a requirement for having space-time metric
  emergent from the pure topology of causal connections, whereas physically homogeneity corresponds
  to the universality of the physical law.  We analyze in detail the case of $1+1$ dimensions. If we
  consider the causal connections as an exchange of classical information, we can establish
  coordinate systems via an Einsteinian protocol, and this leads to a digital version of the Lorentz
  transformations. In a computational analogy, the foliation construction can be regarded as the
  synchronization with a global clock of the calls to independent subroutines (corresponding to the
  causally independent events) in a parallel distributed computation. Thus the Lorentz time-dilation
  emerges as an increased density of leaves within a single \textit{tic-tac} of a clock, whereas
  space-contraction results from the corresponding decrease of density of events per leaf. The
  operational procedure of building up the coordinate system introduces an in-principle
  indistinguishability between neighboring events, resulting in a network that is coarse-grained,
  the thickness of the event being a function of the observer's clock. The illustrated simple
  classical construction can be extended to space dimension greater than one, with the price of
  anisotropy of the maximal speed, due to the Weyl-tiling problem. This issue is cured if the causal
  network is quantum, as e.g. in a quantum cellular automaton, and isotropy is recovered by quantum
  coherence via superposition of causal paths. We thus argue that in a causal network description of
  space-time, the quantum nature of the network is crucial.
\end{abstract}
\maketitle
\section{Introduction}
Our everyday way of looking at space-time as a stage for physical
events conflicts with the requirement of defining all physical
quantities---including space and time---through precise measurement
protocols. This means that we should more properly regard space-time
as emerging from events, instead of pre-existing them. The operational
definition of space-time is defined by the protocol that sets up the
coordinate system. For example, in the Einstein protocol light pulses
are sent back and forth between different locations: at the place
where the signal has been originated, from the arrival time of the
reflected signal one infers both the distance and the time of the
remote event of signal-reflection.  The protocol shows how space-time
is indeed a coherent organization of inferences based on a causal
structure for events. The clock itself is just a sequence of
events---a light pulse bouncing between two mirrors. The closest are
the mirrors, the more precise is the clock, and the more refined is
the coordinate system.

The above reasoning shows that ultimately space and time are defined through pure event-counting,
precisely counting \textit{tic-tacs} of the observer's clock, and we are thus lead to regard space
and time as emergent from the topology of the causal network of events. The events of the network do
not need to be regarded as actual, but can be just potential, and the fabric of space-time is
precisely the network of causal links between them.

The idea of deriving space-time from purely causal structures is not new. Raphael Sorkin started an
independent research line in quantum gravity based on this idea more than two decades ago
\cite{Bombelli-Sorkin_(1987)}. This was motivated by the potentialities of the approach residing in
the natural discreteness of the causal network, which also provides a history-space for a path
integral formulation \cite{Fotini2002,Henson2006}. The possibility of recovering the main features
of the space-time manifold---topology, differentiable structure, and conformal metric---has been
shown, starting from discrete sets of points endowed with a causal partial ordering
\cite{Surya2008}. Along these lines, in an operational context, Lucien Hardy has also formulated a
\textit{causaloid} approach \cite{Hardy-causaloid}, which considers the possibility of a dynamical
treatment of the causal links.

In the causal-set approach of Sorkin events are randomly scattered in order to avoid occurrence of
their sparseness in the boosted frames that would lead to violation of Lorentz invariance. Clearly
the randomness of Sorkin's events should not be regarded in terms of their location on a background,
otherwise we contradict the same idea of space-time emergence. We then need to consider randomnes at
the pure topological level, and this means having random causal connections.  However, regarding the
causal connections as an irreducible description of the physical law, a random topology would then
corresponds to having a random physical law at the most microscopic level (the Planck scale), and
one may argue that a ``random law'' would contradict the same notion of law. Instead, the randomnes
should results from the law itself, e.g. in a quantum cellular automaton, where randomness comes
from quantum nature of the network. The universality of the physical law thus lead us to take the
causal network as topologically homogeneous. Topological homogeneity has then the added bonus that
metric simply emerges from the pure topology by just counting events along the network.  It is
obvious that the discreteness of the network will lead to violation of Lorenz covariance (and the
other space symmetries) at the Planck scale level: however, one must have a theory where covariance
is restored in the large scale limit--the Fermi scale--corresponding to counting huge numbers of
events.

With the above motivations, in this paper we analyze the mechanism of emergence of space-time from
the pure homogeneous topology in $1+1$ dimensions. We present a digital version of the Lorentz
transformations, along with the corresponding digital-analog conversion rule. Upon considering the
causal connections as exchanges of classical information, we can establish coordinate systems via an
Einsteinian protocol, leading to a digital version of the Lorentz transformations. In a
computational analogy first noticed by Leslie Lamport \cite{lamport}, the foliation construction can
be regarded as the synchronization protocol with a global clock of the calls to independent
subroutines (the causally independent events) in a parallel distributed computation.  The boosts are
determined by the relative lengths of the \textit{tic }and \textit{tac} of the clock, and the
Lorentz time-dilation corresponds to an increased number of leaves within a clock \textit{tic-tac},
whereas space-contraction results from the corresponding decreased density of events per leaf, as
first noticed in Ref. \cite{DAriano:QCFT}. 

We will see that the operational procedure of building up the coordinate system introduces an
in-principle indistinguishability between neighboring events, resulting in a network that is
coarse-grained, the thickness of the event being a function of the observer's clock.  The digital
version of the Lorentz transformation is an integer relation which differs from the usual analog
transformation by a multiplicative real constant corresponding exactly to the event thickness. The
composition rule for velocities is independent on such constant, and is the same in both the analog
and the digital versions.  Preliminary results of the present work were already presented in Ref.
\cite{tosini}.

The illustrated simple classical construction can be extended to space dimension greater than one,
but at the price of anisotropy of the maximal speed, due to the Weyl-tiling problem. This issue is
cured if the causal network is quantum, as e.g. in a quantum cellular automaton, and isotropy is
recovered with quantum coherence, corresponding to superposition of causal paths. We will thus argue
that in a causal network description of space-time, the quantum nature of the network is crucial.

\section{Setting up the digital coordinate system}
The first problem to address is which specific lattice should be adopted for the causal network.  In
our convention the causal arrow is directed from the bottom to the top of the network. The dimension
of the emerging space-time corresponds to the graph-dimension of the network, which is the dimension
of the embedding manifold such that all links can be taken as segments of straight line with the
same length.  We will require that the lattice be pure topology (namely with all events equivalent),
corresponding to a locally homogeneous space-time, and with no redundant links. It is then easy to
see that in 1+1 dimensions there are only three possible lattices: the square, the triangular, and
the honeycomb ones. The honeycomb-lattice has two inequivalent types of events (having one input and
two output links and viceversa), and the corresponding ``undressed'' topology---where each couple of
connected inequivalent events are merged into a single event---reduces to the square lattice. The
triangular-lattice, on the other hand, has redundant causal links (the middle vertical ones). We are
thus left with the square-lattice.

\begin{figure}[h]
\includegraphics[width=.4\textwidth]{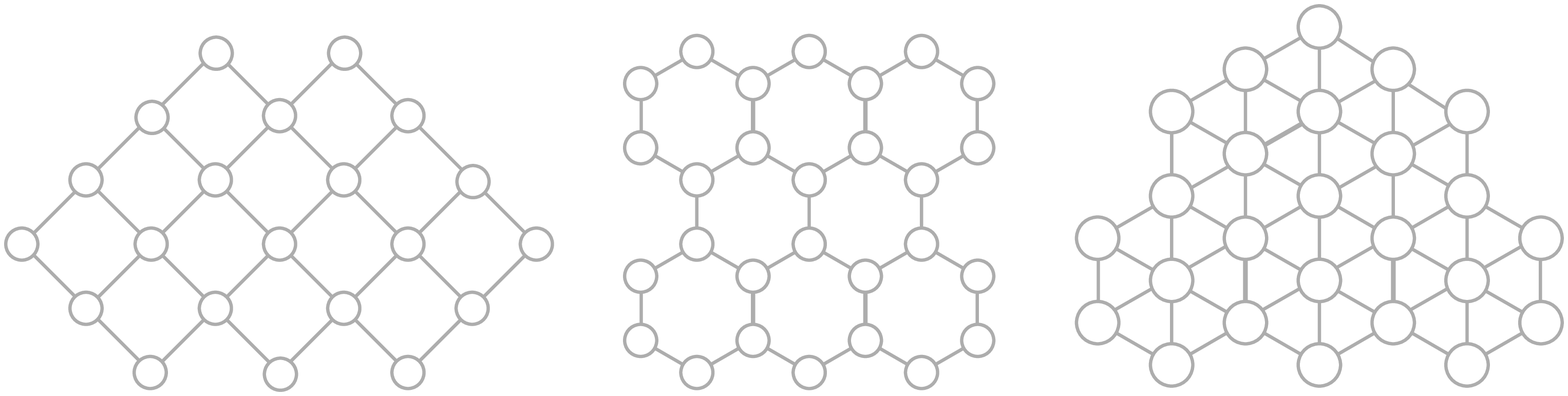} 
  \caption{The three possible homogeneous topologies for causal networks in 1+1 dimensions.}\label{f:1dnets}
\end{figure}

We always assume the network links as oriented according to the causal arrow.  In the square-lattice
network there are thus two types of link: toward the right and toward the left---shortly \textit{r}-link
and \textit{l}-link. Two events are in the same position (for some
boosted reference frame) if they are connected by a path made with a sequence of \textit{r}-links
followed by a sequence of \textit{l}-links. When the two sequences contain the same number of links the
reference is at rest.

A clock is a sequence of causally connected events periodically
oscillating between two positions.  For an Einstein clock the
oscillation (\textit{tic-tac}) is exactly the same couple of sequences
of \textit{l}- and \textit{r-}links identifying events in the same
position. The precision of the clock, namely the minimum amount of
time that it can measure, is the number of links of a complete
\textit{tic-tac}.  The \textit{tic-tac} is indivisible, namely the
sole \textit{tic} (or \textit{tac}) is not a complete measured time
interval, since it involves two different positions.

\begin{figure}[h]
  \includegraphics[width=.23\textwidth]{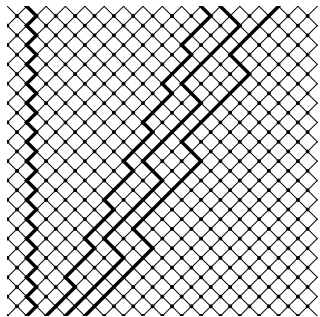}
  \caption{Different clocks on the causal network. The {\em tic-tac} of the clock is represented by
    the two numbers $\alpha$ and $\beta$ (see text). From the left to the right we have the rest-frame clock,
    clock corresponding to $\alpha=\beta=1$, and boosted-frames for $\alpha=3,\beta=1$,
    $\alpha=6,\beta=2$,  and  $\alpha=7,\beta=2$, respectively, corresponding to digital speed
    $v=\frac{1}{3}$, $v=\frac{1}{3}$ and $v=\frac{2}{7}$, respectively. The case $\alpha=6,\beta=2$
    has doubled imprecision, compared to the case $\alpha=3,\beta=1$. 
    \label{f:clocks}}
\end{figure}

In the following we will call \textit{light signals} those sequences of events that are connected
only by \textit{r}-links or by \textit{l-}links, namely making segments at 45 degrees with the
horizontal in the network. Their ``speed'' is equal to ``one event-per-step'', and is the maximum
speed allowed by the causality of the network, since connecting events along a line making an angle
smaller than $45^\mathit{o}$ with the horizontal would require to follow some causal connections in
the backward direction from the output to the input. In this way in a homogeneous causal network
suffices to guarantee a bound for the speed of information flow.

\begin{figure}[h]
  \includegraphics[width=.23\textwidth]{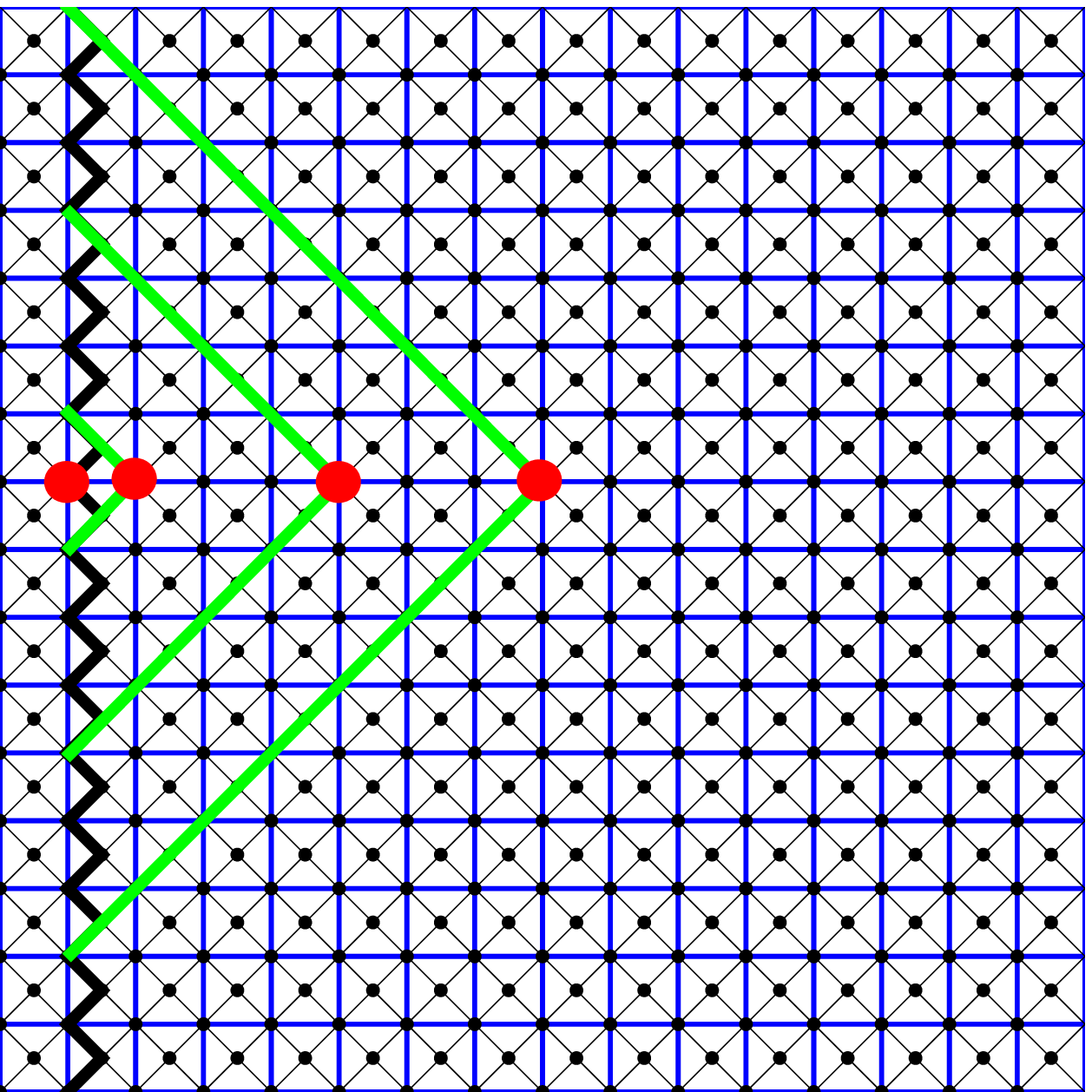}
  \includegraphics[width=.23\textwidth]{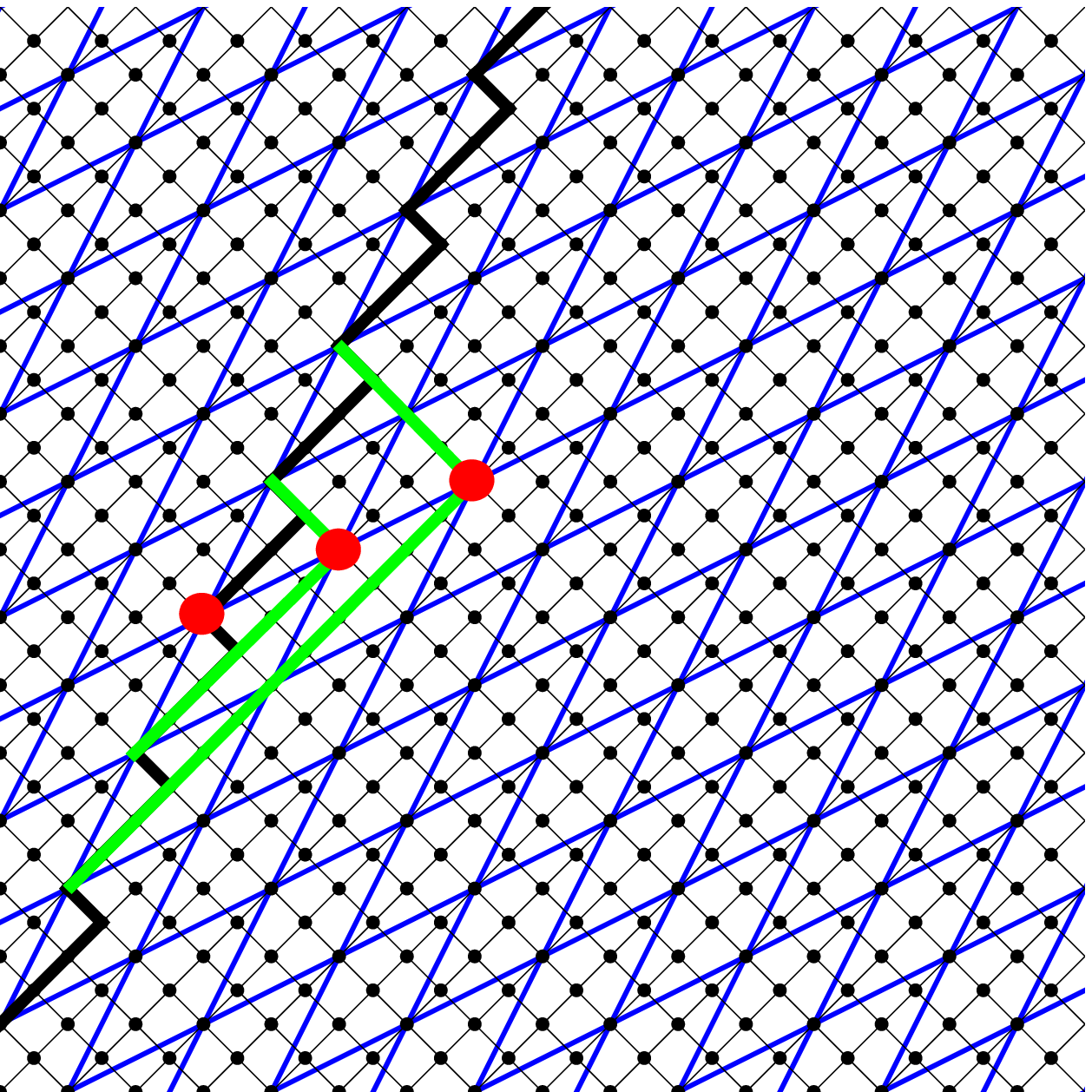}
  \caption{Illustration of the Einstein protocol for establishing a reference frame.  {\bf Left
      figure:} Rest-frame. The blue lines represent the reference frame established using the clock with
    $\alpha=\beta=1$ (see Fig. \ref{f:clocks}). The green lines represent light signals bouncing
    between the clock and four particular events in the network. These events are synchronous, since
    the intermediate time between the sending and the receiving event on the clock is the same for
    all of them. They lie on the same leaf of the foliation, but at different position, $0,1,4,7$, respectively: the spatial
    coordinate is obtained by counting the \emph{tic-tacs} between the the sending and the receiving
    event ($0,2,8,14$,respectively) divided by two. {\bf Right figure:} Boosted reference frame
    (blue lines) for $\alpha=3,\beta=1$, built up using the same protocol as in the felt figure.
    \label{f:clocks2}}
\end{figure}

In the following we will take the clock \textit{tic-tac} made with
$\alpha$ \textit{r}-links followed by $\beta$ \textit{l}-links (see
Fig, \ref{f:clocks}). Any clock allows to introduce a reference frame
$\R$ which is just a foliation of the network built up using the
Einstein protocol.  From the start of the clock \textit{tic-tac} a
light signal is sent to an event in a different position and then
received back at the clock. The intermediate time between the sending
and the receiving event is taken as synchronous with the event at the
turning point, and the number of \textit{tic-tacs} divided by two is
taken as the distance from the turning point and the clock
conventionally located at the beginning of the \textit{tic-tac}. In
this way we build the foliation corresponding to a given clock. A set
of synchronous events identifies a leaf of the foliation. In Fig.
\ref{f:clocks2} the Einstein protocol is illustrated in two particular
reference frames. The figure on the left corresponds to the
rest-frame, with the blue lines depicting the coordinate system
established using the clock with $\alpha=\beta=1$ (see
Fig. \ref{f:clocks}). The green lines represent light signals bouncing
between the clock and four particular events in the network. These
events are synchronous, since the intermediate time between the
sending and the receiving event on the clock is the same for all of
them. They lie on the same leaf of the foliation, but at different
position, $0,1,4,7$, respectively: the spatial coordinate is obtained
by counting the \emph{tic-tacs} between the the sending and the
receiving event divided by two. The right figure represents a boosted
frame for $\alpha=3,\beta=1$, built up using the same protocol as in
the left figure.

\begin{figure}[h]
  \includegraphics[width=.28\textwidth]{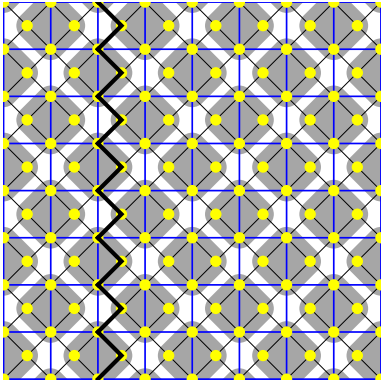}
  \caption{Coarse-graining in the rest-frame. The coarse-graining due to boosts is trivial (depicted
    in yellow, containing just one event), whereas the coarse-graining associated to the intrinsic
    imprecision of the clock is not trivial (depicted in gray, containing four
    events).\label{f:foliation11}}
\end{figure}

\begin{figure}[h]
  \includegraphics[width=.28\textwidth]{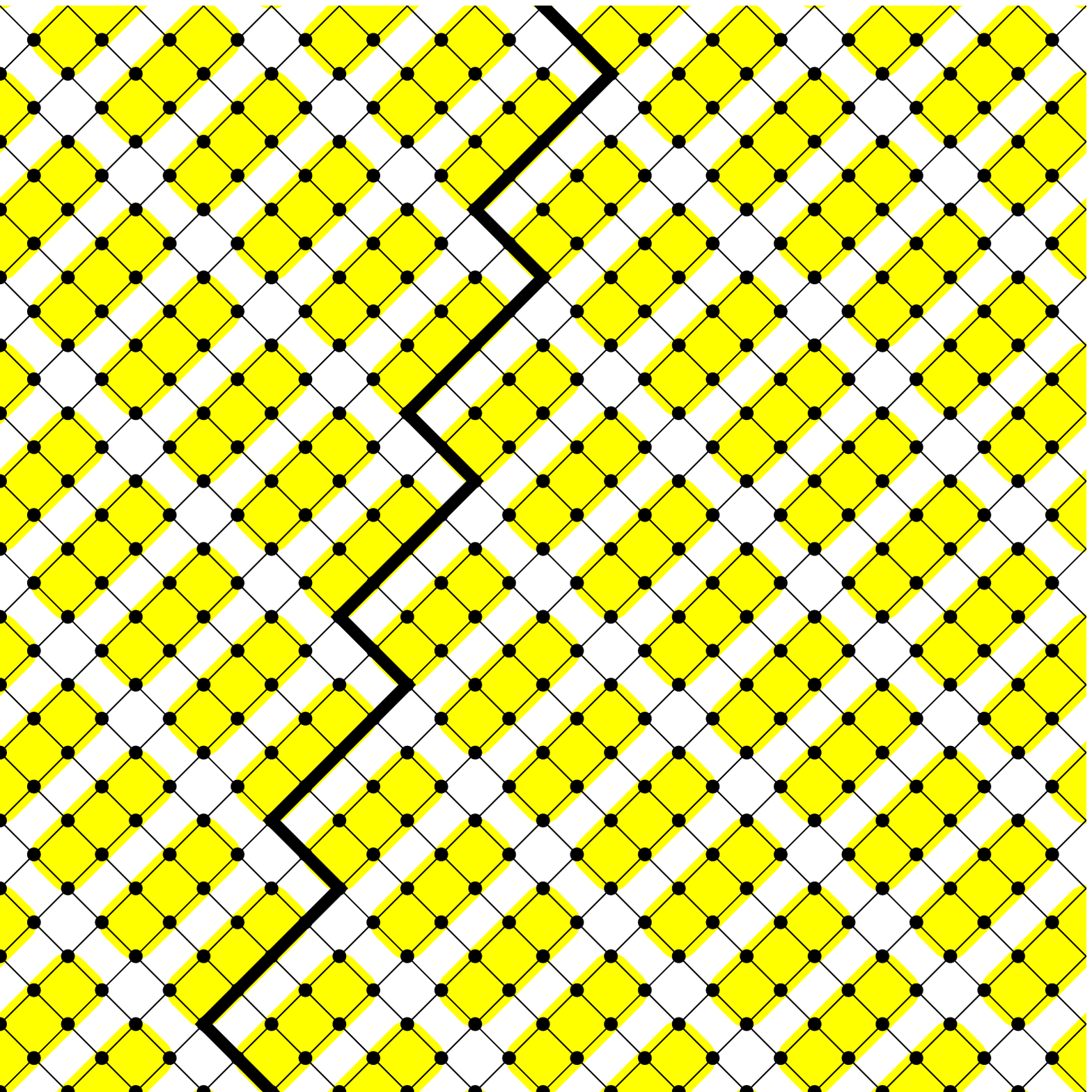}\vskip .1cm
  \includegraphics[width=.28\textwidth]{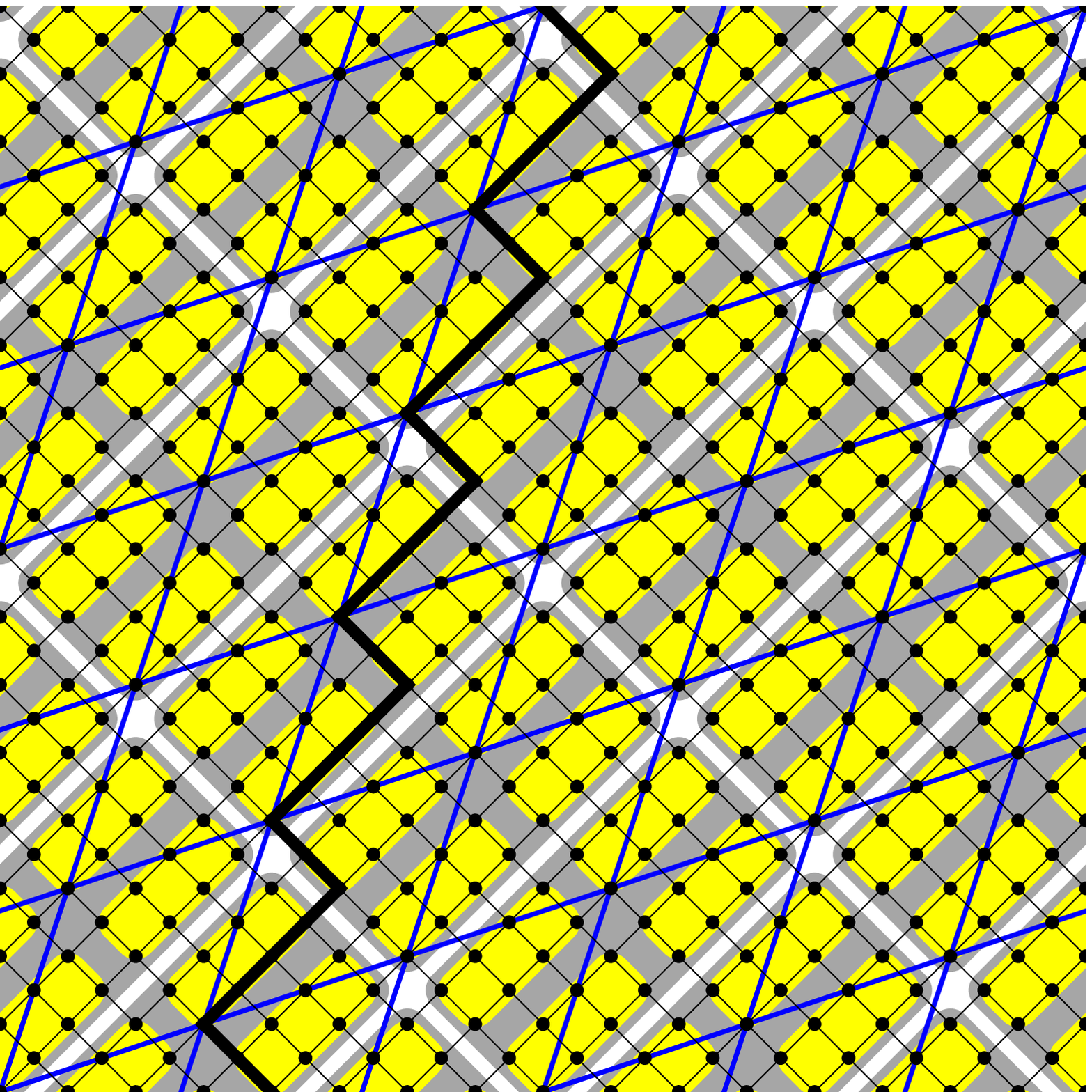} \vskip .1cm
  \includegraphics[width=.28\textwidth]{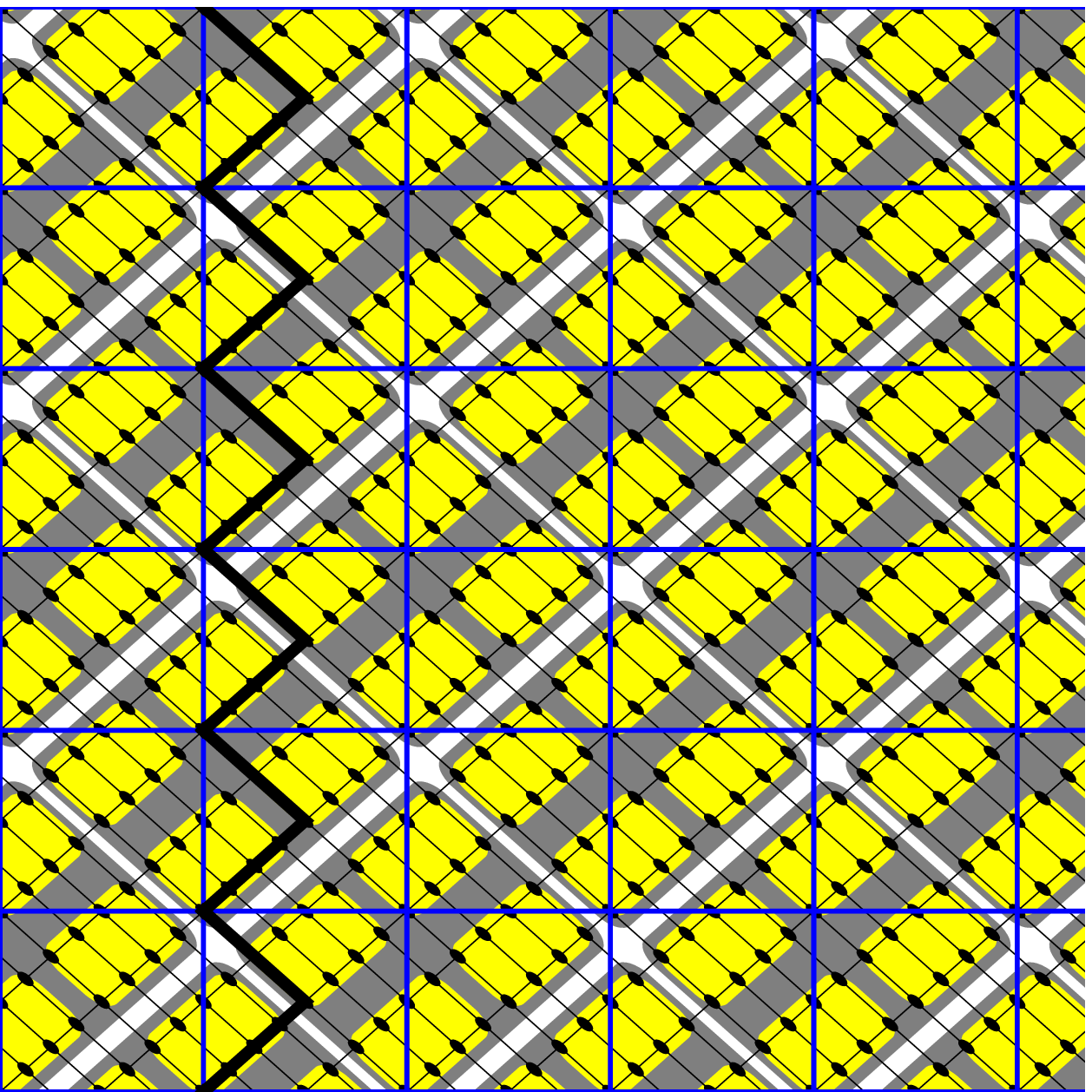}
  \caption{Illustration of the coarse-graining procedure due to boost (in yellow) and to intrinsic
    imprecision of the clock (in gray). {\bf Top figure:} events along the \textit{tic} and events along the
    \textit{tac} are identified in the boosted frame. Then events are merged into minimal sets (in
    yellow) so that the topology is left invariant (the merged events are again events of a square-lattice
    network). {\bf Center figure:} the coarse graining associated to the intrinsic imprecision is added
    in gray. {\bf Bottom figure:} the circuit is stretched so to have all synchronous events on
    horizontal lines, and events located in the same position on vertical lines.
    \label{f:foliation_coarse}}
\end{figure}

Due to indivisibility of the \textit{tic-tac}, we see that there are
indiscernible events, for which the synchronisation occurs in the
middle of the \textit{tic-tac}. We are thus led to identify events,
and merge them into thicker coarse-grained events. This is done as
follows. We identify the events along the \textit{tic} and those along
the \textit{tac} so that the \textit{tic-tac} is always regarded as
the bouncing between two next neighbour events. Then we merge events
into minimal sets so that the topology is left the invariant (see
figures \ref{f:foliation11} and \ref{f:foliation_coarse}). We can
distinguish between two different kinds of coarse-graining: one due to
the boosting (in yellow in the figures), and one due to intrinsic
imprecision of the clock (in gray).  The difference between the two is
clarified in Fig. \ref{f:foliation_coarse}.  In the top figure events
along the \textit{tic} and events along the \textit{tac} are
identified in the boosted frame. Then events are merged into minimal
sets (in yellow) so that the topology is left invariant (the merged
events are again events of a square-lattice network). In the central
figure the coarse graining associated to the intrinsic imprecision is
added in gray, and finally, in the bottom figure the circuit is
stretched so to have all synchronous events on horizontal lines, and
events located in the same position on vertical lines. Notice that in
the special case of the rest-frame, see Fig. \ref{f:foliation11},
the coarse-graining is just due to the intrinsic imprecision of the
clock.

\section{The digital Lorentz transformations}
The velocity of the boosted frame can be easily written in terms of the $\alpha$ and $\beta$ of the
{\em tic-tac} of the clock, by simply evaluating the ratio of the distances in space and time between the
two ending points of the {\em tic-tac}, namely
\begin{equation}\label{e:speed1}
v=\frac{\alpha-\beta}{\alpha+\beta}.
\end{equation}
Thanks to invariance of topology under the boost coarse-graining, the
above identity holds also for the motion relative to any boosted
frame, whence, upon defining with
$\alpha_{12}=\alpha_2/\alpha_1\in\mathbb Q$ ($\mathbb Q$ denoting
rational numbers) and $\beta_{12}=\beta_2/\beta_1\in\mathbb Q$ for
frames $\R_1$ and $\R_2$, and by $v_{12}\in\mathbb Q$ the relative
velocity of frame $\R_2$ with respect to frame $\R_1$, one has
\begin{equation}\label{e:speed12}
v_{12}=\frac{\alpha_{12}-\beta_{12}}{\alpha_{12}+\beta_{12}}.
\end{equation}
Now, by using the trivial identities $\alpha_{13}=\alpha_{12}\alpha_{23}$ and
$\beta_{13}=\beta_{12}\beta_{23}$ one has
\begin{equation}\label{e:network_vel_comp}
  v_{13}=\frac{\a^{12}\a^{23}-\b^{12}\b^{23}}
  {\a^{12}\a^{23}+\b^{12}\b^{23}},
\end{equation}
which by simple algebraic manipulations immediatley gives
\begin{equation}
  v_{13}=\frac{\left(\frac{\a^{12}-\b^{12}}{\a^{12}+\b^{12}}\right)+
    \left(\frac{\a^{23}-\b^{23}}{\a^{23}+\b^{23}}\right)}
  {1+\left(\frac{\a^{12}-\b^{12}}{\a^{12}+\b^{12}}\right)
    \left(\frac{\a^{23}-\b^{23}}{\a^{23}+\b^{23}}\right)}=
  \frac{v_{12}+v_{23}}{1+v_{12}v_{23}}.
\end{equation}
The last identity is the composition rule of parallel velocities (the
only possibility in $1+1$ dimension) in special relativity.

\begin{figure}[h]
  \includegraphics[width=.235\textwidth]{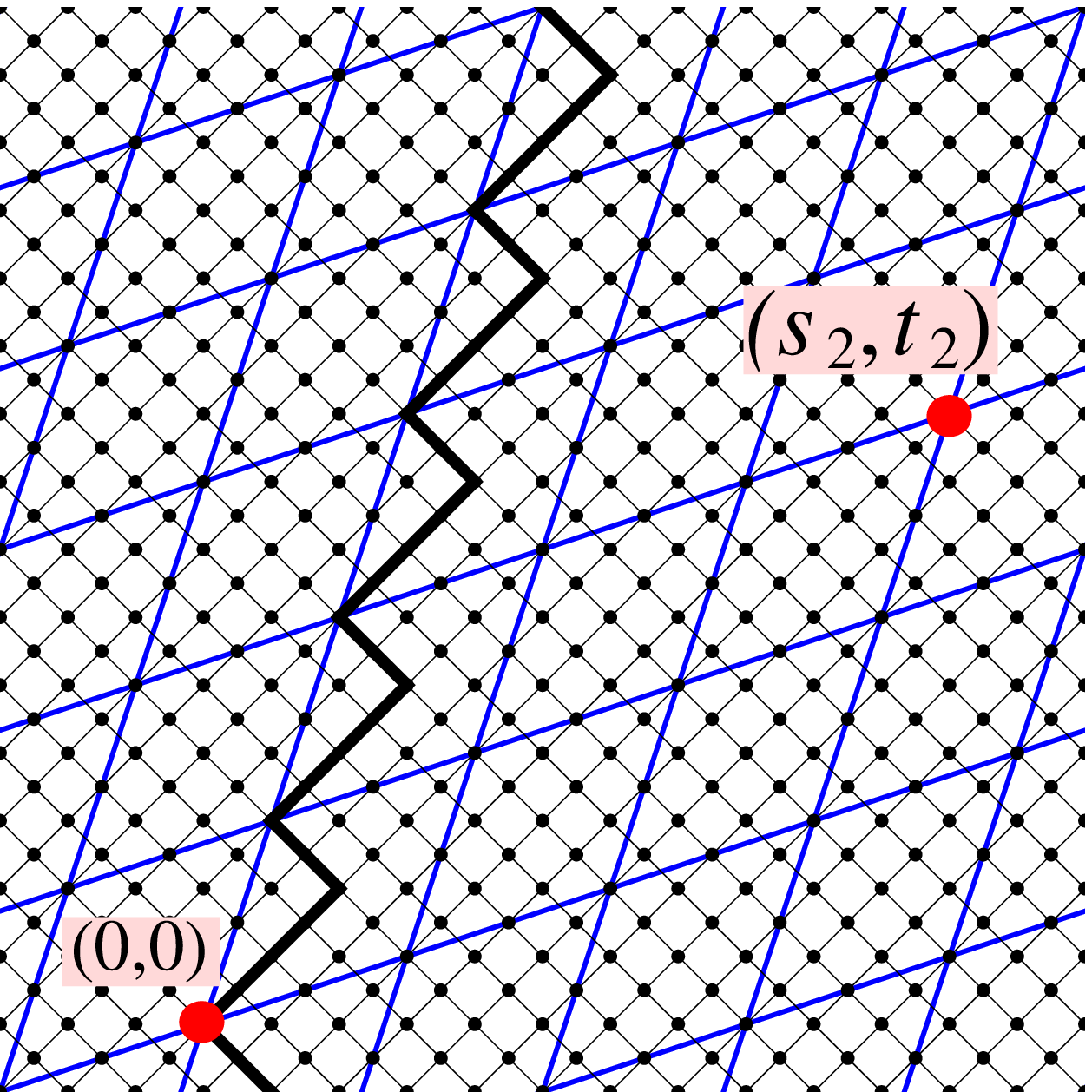}
  \includegraphics[width=.235\textwidth]{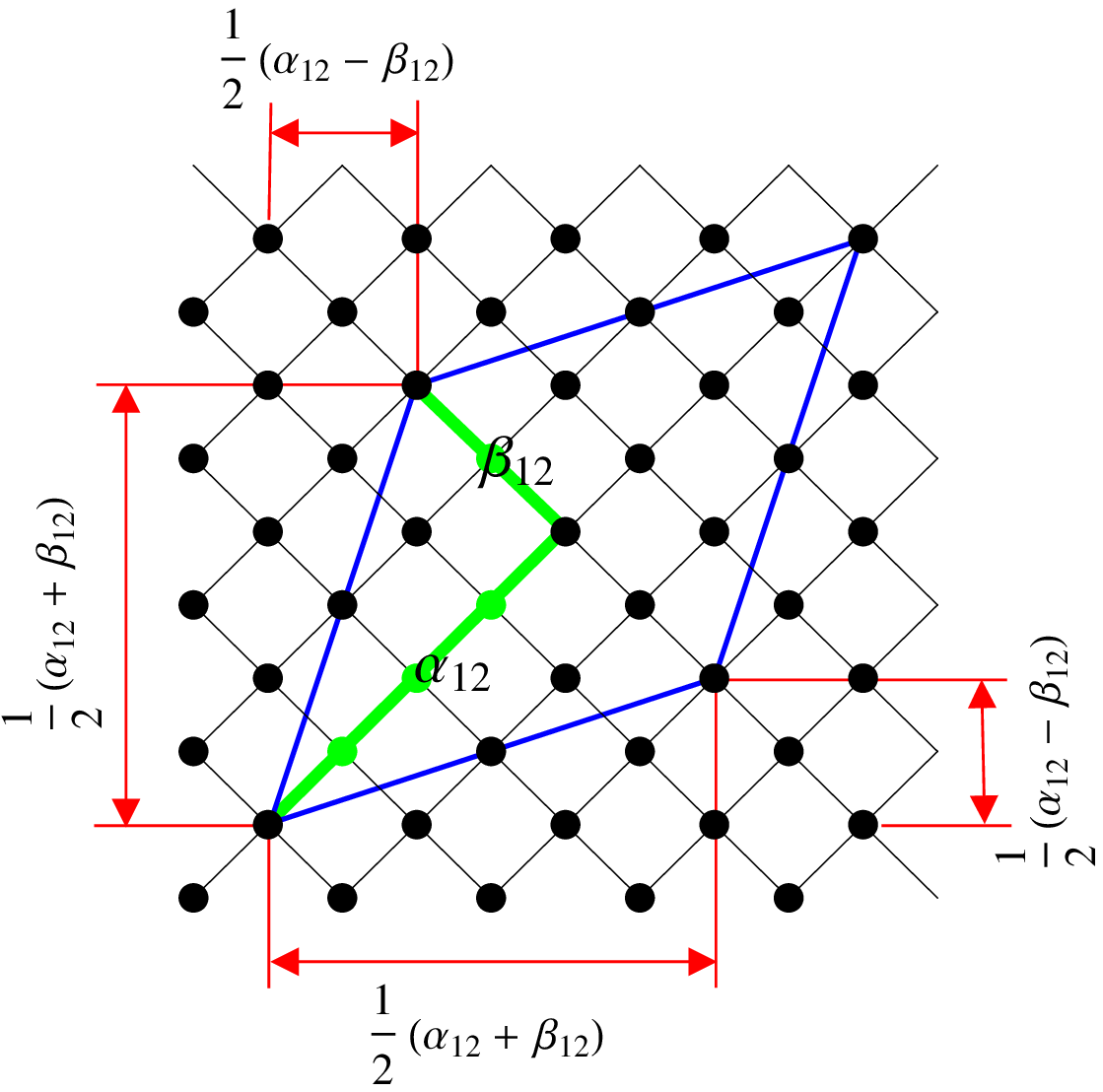}
  \caption{Illustration of the derivation of Eqs. (\ref{e:Lorentztrans1}) and
    (\ref{e:Lorentztrans2}), leading to the digital version of the Lorentz transformations
    (\ref{ttrans1}) and (\ref{ttrans2}). {\bf Left
    figure:} the reference frame $\R_1$ with $\alpha_1=\beta_1=1$ is represented by the tyny network
  in black, whereas the coarser network in blue represents the boosted reference frame $\R_2$ with $\alpha_2=4$,
    $\beta_2=2$. According to Eq. (\ref{e:speed12}) the relative velocity of $\R_2$ with respect to
    $\R_1$ is $v_{12}=1/3$. In  order to connect the coordinate systems in the two frames we have
    chosen the same origin $(0,0)$ on both $\R_2$ and $\R_1$. The generic event on $\R_2$ has coordinates
    $(s_2,t_2)=(3,2)$ and $(s_1,t_1)=(10,9)$ in the two frames, respectively. {\bf Right figure:}
    a spatial step in $\R_2$ corresponds to $(\alpha_{12}+\alpha_{12})/2$ space and
    $(\alpha_{12}-\alpha_{12})/2$ time steps in $\R_1$. In the same way to a time step in $\R_2$ corresponds
    to $(\alpha_{12}-\alpha_{12})/2$ space and $(\alpha_{12}+\alpha_{12})/2$ time steps in $\R_1$. This
    correspondence allows to determine the coordinates of a given event in the 
    frame $(s_1,t_1)$  in terms of its coordinates $(s_2,t_2)$ in the frame $\R_2$.
    The resulting transformations are in Eqs. (\ref{e:Lorentztrans1}) and
    (\ref{e:Lorentztrans2}).
    \label{f:Lorentztrans}}
\end{figure}
 
Now we use the Einstein protocol to construct the boosted coordinated system with respect to the
rest-frame along with the relative coordinate systems between any couple of boosted frames. We will
now see that the coordinates of an event transform from the frame $\R_2$ to the frame $\R_1$ as
follows
\begin{align}
\label{e:Lorentztrans1}
    s_1=\tfrac{1}{2}(\a_{12}+\b_{12})s_2+\tfrac{1}{2}(\a_{12}-\b_{12})t_2,\\
\label{e:Lorentztrans2}
    t_1=\tfrac{1}{2}(\a_{12}+\b_{12})t_2+\tfrac{1}{2}(\a_{12}-\b_{12})s_2.
\end{align}
In fact, from a simple inspection of Figure \ref{f:Lorentztrans} one can check Eqs.
(\ref{e:Lorentztrans1}) and (\ref{e:Lorentztrans2}) with the frame $\R_1$ as the rest frame.  In the
left figure the reference frame $\R_1$ with $\alpha_1=\beta_1=1$ is represented by the tyny network
in black, whereas the coarser network in blue represents the boosted reference frame $\R_2$ with
$\alpha_2=4$, $\beta_2=2$. According to Eq. (\ref{e:speed12}) the relative velocity of $\R_2$ with
respect to $\R_1$ is $v_{12}=1/3$. In order to connect the coordinate systems in the two frames we
have chosen the same origin $(0,0)$ on both $\R_2$ and $\R_1$. The generic event on $\R_2\cap\R_1$
has coordinates $(s_2,t_2)=(3,2)$ and $(s_1,t_1)=(10,9)$ in the two frames, respectively. In the
figure on the right one can see that a spatial step in $\R_2$ corresponds to
$(\alpha_{12}+\alpha_{12})/2$ space and $(\alpha_{12}-\alpha_{12})/2$ time steps in $\R_1$. In the
same way a time step in $\R_2$ corresponds to $(\alpha_{12}-\alpha_{12})/2$ space and
$(\alpha_{12}+\alpha_{12})/2$ time steps in $\R_1$. This correspondence allows to determine the
coordinates $(s_1,t_1)$ of a given event in the frame $\R_1$ in terms of its coordinates $(s_2,t_2)$
in the frame $\R_2$.  The resulting transformations are in Eqs. (\ref{e:Lorentztrans1}) and
(\ref{e:Lorentztrans2}).  Invariance of topology with boost, guarantees that they also hold between
any couple of boosted frames. By elementary manipulation Eqs. (\ref{e:Lorentztrans1}),
(\ref{e:Lorentztrans2}) can be written in the more customary way
\begin{align}
  s_1=\tfrac{1}{2}(\a_{12}+\b_{12}) \left( s_2+v_{12}t_2 \right),\label{ttrans1}\\
  t_1=\tfrac{1}{2}(\a_{12}+\b_{12}) \left( t_2+v_{12}s_2 \right).\label{ttrans2}
\end{align}
Upon defining the following constant depending on the clocks of the
two frames
\begin{equation}\label{e:chi}
\chi_{12}:=\sqrt{\a_{12}\b_{12}},
\end{equation}
and using the identity
\begin{equation}\label{e:gamma}
  \tfrac{1}{2}(\a_{12}+\b_{12})=\frac{\chi_{12}}{\sqrt{1-v_{12}^2}},
\end{equation}
we obtain the {\em digital} Lorentz transformations
\begin{equation}\label{digitalLorentz}
 s_1=\chi_{12}\frac{s_2+v_{12}t_2}{\sqrt{1-v_{12}^2}},\quad
 t_1=\chi_{12}\frac{t_2+v_{12}s_2}{\sqrt{1-v_{12}^2}}.
\end{equation}

\begin{figure}[h]
  \includegraphics[width=.4\textwidth]{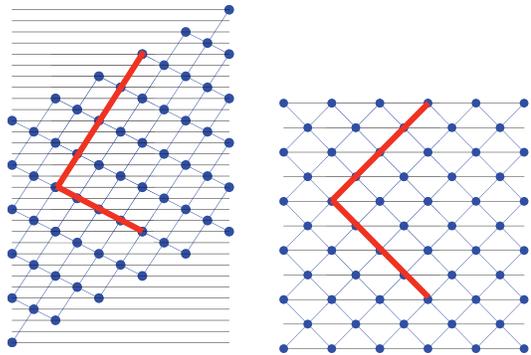}
  \caption{The mechanism for the digital Lorentz time-dilation and space-contraction given in Ref.
    \cite{DAriano:QCFT}, here for a boost with $v=\frac{1}{2}$, corresponding to a digital
    time-dilation by a factor 2 (analog factor $2/\sqrt3$) and space-contraction by a factor $1/2$
    (compare with the same factors in Eqs.  (\ref{digitalLorentz})).\label{f:Lorentz}}
\end{figure}
Eqs. (\ref{digitalLorentz}) differ from the usual {\em analog} Lorentz
transformations by the multiplicative factor $\chi_{12}$, which is
logically required to make the transformations rational, compensating
the irrationality of the boost factor $\sqrt{1-v_{12}^2}$. The
digital-analog conversion is thus just a rescaling of both space and
time coordinates by the factor $(\alpha\beta)^{\frac{1}{2}}$ depending
on the boost, which is exactly the square-root of the volume of the
coarse-grained event measured as the number of rest-frame events that
it contains.  Such event volume also affects the Lorentz
space-contraction and time-dilation factor, which in the digital case
is given by $\frac{1}{2}(\alpha_{12}+\beta_{12})$, whereas in the
analog case is rescaled by the ratio of event volumes, leading to
$\frac{1}{2\sqrt{\alpha_{12}\beta_{12}}}(\alpha_{12}+\beta_{12})$. Thus,
for example, for $\alpha_{12}=1$ and $\beta_{12}=3$ corresponding to
$v_{12}=1/2$ the digital factor is $2$ whereas the analog one is
$2/\sqrt3 $. The digital factor agrees with that of the Lorentz
time-dilation and space-contraction mechanism of Ref.
\cite{DAriano:QCFT}, given in terms of increased density of leaves and
corresponding decreased density of events per leaf, as illustrated in
Fig.  \ref{f:Lorentz}.

\section{Conclusions and discussion}
We have analyzed the mechanism of emergence of space-time from
homogeneous topology in $1+1$ dimensions, deriving the digital version
of the Lorentz transformations along with the corresponding
digital-analog conversion rule. The homogeneity of topology physically
represents the universality of the physical law (it is worth
mentioning that such law is stripped of the conventionality of space
and time homogeneity: see e.g. Ref. \cite{Malament}). We have built
the digital coordinate system using the Einstein's protocol, with
signals sent back-and-forth to events from an observer's clock. We
found that the procedure introduces an in-principle
indistinguishability between neighbouring events, due to the limited
precision of the clock, resulting in a network that is coarse-grained,
with the event thickness also depending on the boost. The digital
version of the Lorentz transformation is an integer relation which
differs from the usual analog transformation by a multiplicative real
constant corresponding to the event thickness.

\begin{figure}[h]
\includegraphics[width=.3\textwidth]{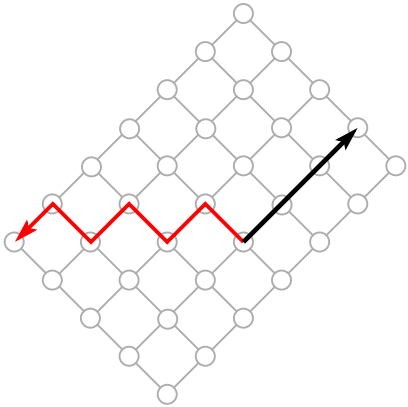}
\caption{Square $(1+2)$-dimensional computational network: view of a
  leaf in the rest frame.  Information must zig-zag to flow at the
  maximal speed in diagonal direction. This leads to a slow-down of a
  $\sqrt{2}$ factor of the analog speed compared to cubic axis
  direction.}\label{f:Pyth}
\end{figure}

\begin{figure}[h]
\includegraphics[width=.45\textwidth]{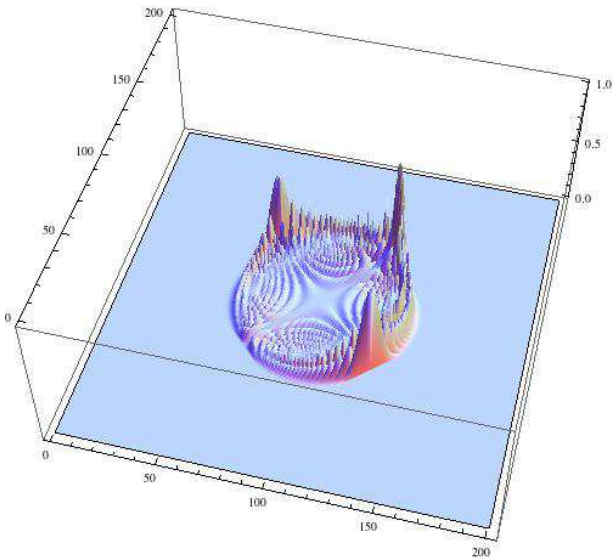} 
\caption{Evolution for 60 steps of the probability distribution of
  finding a particle or antiparticle in a two-dimensional quantum Weyl
  automaton of the kind of Bialynicki-Birula \cite{Bialynicki-Birula}
  on a square-lattice. One can see how the propagation speed is
  isotropic after few steps.}\label{f:automaton}
\end{figure}

The present purely classical cinematical construction does not straightforwardly extend from one
dimension to larger dimensions, due to the Weyl-tiling issue, namely that continuum geometry cannot
simply emerge from counting sites on a discrete lattice, since e.g. in a square tiling one counts
the same number of tiles along a side and along the diagonal of a square \cite{Weyl}. Thus, for
example, as shown in Fig. \ref{f:Pyth}, in a causal network shaped as a square-lattice the fastest
speed would be along the cubic axes, whereas along diagonals information should zig-zag, resulting
in a slowdown by a factor $\sqrt{2}$ (or even $\sqrt{3}$ in three dimensions). Indeed, a general
theorem of Tobias Fritz \cite{tobias} shows that the polytope of points that can be reached in no
more than $N$ links in a periodic graph does not approach a circle for large $N$. Since the polytope
has necessarily distinguished directions, this means that there is no periodic graph for which this
velocity set is isotropic.  This result represents a no-go theorem for the emergence of an isotropic
space from a discrete homogeneous causal network representing a {\em classical} information flow.

The situation, however, is completely different if one considers the possibility that information
can flow in a superposition of paths, along the network, as in a quantum cellular automaton,
corresponding to a homogeneous quantum computational network. In Fig. \ref{f:automaton} a concrete
example of evolution is given for a two-dimensional quantum Weyl automaton of the kind of
Bialynicki-Birula \cite{Bialynicki-Birula} on a square-lattice. One can see that the maximum
propagation speed is isotropic after just few steps. In a similar way full Lorentz covariance is
expected to be restored in the same limit of infinitely many events--a kind of thermodynamic limit
bringing the automaton to the Fermi scale.

\bigskip
\section*{Acknowledgements}
GMD acknowledges interesting discussions with Raphael Sorkin, Seth Lloyd, and Tobias Fritz. This
work has been partially supported by PRIN 2008.


\begin{thebibliography}{15}
\bibitem{Bombelli-Sorkin_(1987)} L.~Bombelli, J.~H.~Lee, D.~Meyer, and R.~Sorkin, Phys. Rev. Lett {\bf 59}, 521 (1987).
\bibitem{Fotini2002} F.~Markopoulou, gr-qc/0210086 (2002).
\bibitem{Henson2006} J.~Henson, in \emph{Approaches to Quantum Gravity: Towards a New Understanding
    of Space and Time}, Ed. D.~Oriti (Cambridge University Press, Cambridge UK, 2006) (also
  gr-qc/0601121).
\bibitem{Surya2008} S.~Surya, Theor. Comp. Sc. {\bf 405} 188 (2008).
\bibitem{Hardy-causaloid} L. Hardy, J. Phys. A {\bf 40} 3081 (2007).
\bibitem{lamport} L. Lamport, Comm. ACM, {\bf 21} 558 (1978).
\bibitem{DAriano:QCFT} G.~M.~D'Ariano, in CP1232 {\em Quantum Theory: Reconsideration of
    Foundations, 5} ed. by A.~Y. Khrennikov, (AIP, Melville, New York, 2010), pg. 3 (also arXiv:
  1001.1088).
\bibitem{tosini}  G. M. D'Ariano and A. Tosini,  arXiv:1008.4805 (2010)
\bibitem{Malament} D. Malament, No{\^u}s {\bf 11} 293 (1977).
\bibitem{Weyl} H. Weyl, {\em Philosophy of Mathematics and Natural Sciences}, Princeton University
\bibitem{tobias} T. Fritz, {\em Velocity Polytopes of Periodic Graphs}, draft (2011).
  Press, (Princeton 1949).
\bibitem{Bialynicki-Birula} I. Bialynicki-Birula, Phys. Rev. D {\bf 49} 6920 (1994).
\end{thebibliography}
\end{document}